\documentclass[draft]{agujournal2019}
\usepackage{url} 
\usepackage{amsmath,amssymb}
\usepackage{lineno}
\usepackage{siunitx}  
\usepackage{ragged2e} %
\justifying
\draftfalse

\journalname{Geophysical Research Letters}

\begin{document}

%
%


\title{Supersaturation in the Wake of a Precipitating Hydrometeor and its Impact on Aerosol Activation}

%
%




\authors{Taraprasad Bhowmick\affil{1,2}, Yong Wang\affil{2}, Michele Iovieno\affil{3}, Gholamhossein Bagheri\affil{2}, and Eberhard Bodenschatz\affil{2,4,5}}
\affiliation{1}{Department of Applied Science and Technology, Politecnico di Torino, Torino, Italy}
\affiliation{2}{Laboratory for Fluid Physics, Pattern Formation and Biocomplexity, Max Planck Institute for Dynamics and Self-Organization, G\"ottingen, Germany}
\affiliation{3}{Department of Mechanical and Aerospace Engineering, Politecnico di Torino, Torino, Italy}
\affiliation{4}{Institute for Dynamics of Complex Systems, University of G\"ottingen, G\"ottingen, Germany}
\affiliation{5}{Laboratory of Solid State Physics, Cornell University, Ithaca, NY, USA}






\correspondingauthor{Yong Wang}{yong.wang@ds.mpg.de}
\correspondingauthor{Gholamhossein Bagheri}{gholamhossein.bagheri@ds.mpg.de}




\begin{keypoints}
\item This study shows how wake-induced supersaturation in clouds activates aerosols in an equal level of other secondary production processes.
\item The parameter space for wake-induced supersaturation behind precipitating spherical hydrometeors is detailed.
\item It is described how lucky aerosols are activated in such supersaturated wake of the precipitating hydrometeors.
\end{keypoints}

%
%

%
%
\graphicspath{{./Figures/}}

\newcommand{\dr}{{\rm d}}
\makeatletter
\newcommand\abs{\@ifstar\lr@abs\n@abs}
\newcommand\lr@abs[1]{\left|#1\right|}
\newcommand\n@abs[2][]{\mathopen{#1|}#2\mathclose{#1|}}
\makeatother

\newcommand{\umiss}[1]{\;{\mbox{\rm #1}}}
\newcommand{\umis}[1]{{\mbox{\rm #1}}}

\newcommand{\emme}{{\cal M}}
\newcommand{\au}{\text{\textit{au}}}
\newcommand{\vs}{\text{\textit{vs}}}
\newcommand{\de}{{\rm d}}

\DeclareRobustCommand{\vett}[1]{
  \ifcat#1\relax
    \boldsymbol{#1}
  \else
    \mathbf{#1}
  \fi}
\newcommand{\tvett}[1]{
\mbox{$\tilde{\mathbf #1}$}
}
\newcommand{\pvett}[1]{
\mbox{$\dot{\mathbf #1}$}
}
\newcommand{\media}[1]{\langle #1 \rangle}



\begin{abstract}
The secondary activation of aerosols impacts the life cycle of a cloud. A detailed understanding is necessary for reliable climate prediction.  Recent laboratory experiments demonstrate that aerosols can be activated in the wake of precipitating hydrometeors. However, many quantitative aspects of this wake-induced activation remain unclear. Here, we report a detailed numerical investigation of the activation potential of wake-induced supersaturation. By Lagrangian tracking of aerosols we show that a significant fraction of aerosols are activated in the supersaturated wake. These `lucky aerosols' are entrained in the wake's vortices and reside in the supersaturated environment sufficiently long to be activated. Our analyses show that wake-induced activation can contribute at a level similar to other well known secondary production processes.
\end{abstract}

\section*{Plain Language Summary}

We numerically investigate how new water droplets or ice particles are formed within a cloud. Out of several proposed physical processes for droplet generation, recent experimental studies have shown that a large droplet can nucleate aerosols in the wake behind it when falling under gravity. We present a detailed analysis of various physical factors that lead to an excess of water vapor behind the hydrometeors (e.g., droplets, sleet or hail) and investigate the effectiveness of this process on activation of aerosols to create new cloud particles.

%
%

%


%
%
%
%

\section{Introduction}

The dynamics of atmospheric clouds remains a major source of uncertainty in weather and climate models \cite{stevens2013climate} due to the interplay of many physical processes over a wide range of scales \cite{Bodenschatz2010}.
Especially the activation of aerosols and species therein controls the lifetime of a cloud \cite{Kreidenweis2019} in which fractions of cloud condensation nuclei (CCN) and ice nucleating particles (INP) develop into new hydrometeors \cite{Baker1997}. Physical processes contributing to the activation \cite{field2017} within a mature cloud can not explain the observed discrepancies between the measured activation and the observed hydrometeor population, which is several orders of magnitude higher than expected \cite{Pruppacher2010,Huang2017}.
One possible explanation of this riddle might be the recently discovered  wake-induced supersaturation and activation of aerosols behind large precipitating hydrometeors, i.e. heterogeneous wake-induced nucleation \cite{Prabhakaran2017,Chouippe2019,Prabhakaran2020}.
The experimental investigation by \citeA{Prabhakaran2017} of falling drops (diameter of $\mathcal{O}(1)$ mm) in near critical point conditions of pressurized sulfur-hexafluoride showed evidences of homogeneous nucleation in the wake. 
\citeA{Prabhakaran2020} conducted a follow-up experiment on heterogeneous nucleation using sodium chloride and silver iodide aerosols under atmosphere-like conditions.
Warm droplets with a diameter of $\sim 2$ mm were able to induce the activation of ice aerosols in their wake when precipitating through a subsaturated colder environment. Earlier, a numerical analysis of supersaturation in the wake of a warmer hydrometeor moving through various colder environments was performed by \citeA{Chouippe2019}.
Their work confirms the existence of a supersaturated region in the wake of a hydrometeor that settles through a colder saturated environment. 
The maximum supersaturation observed in the wake was higher the larger the  temperature difference between the hydrometeor and the ambient was.
In a more recent study, \citeA{Chouippe2020} extended their earlier work \citeA{Chouippe2019} and explicitly estimated the influence of wake supersaturation on the ice enhancement factor using a model based on a power law dependence of the local supersaturation \cite{Huffman1973,Baker1991} and concluded that the local ice nucleation enhancement alone cannot produce a sufficient number of activated ice nuclei to solve the observed number discrepancy. Although the development of supersaturation was studied numerically, the direct calculation of nucleation remained quite difficult due to the large number of parameters that include size distribution \cite{Dusek2006}, number concentration \cite{Baker1997}, chemical composition \cite{Curtius2009,DeMott2018}, porosity or solubility \cite{Kanji2017} of aerosols etc. The complexity in the nucleation of  aerosols is further complicated in mixed-phase clouds containing both water and ice phase hydrometeors.
Ice nucleation through deposition and condensation freezing can occur on an aerosol during supersaturation in the ice phase at sub-zero temperatures \cite{Mayers1992}.
Activation of aerosols by immersion freezing on a CCN or by contact freezing on a supercooled water drop can also be observed \cite{Kanji2017}.

The above mentioned studies elucidated some aspects of wake supersaturation and aerosol initiation. 
In this letter, we present a comprehensive numerical study covering the parameter space relevant for atmospheric situations. 
We quantify the influence of ambient humidity and ambient/hydrometeor temperatures on the supersaturation within the wake for different sizes and phases of spherical hydrometeors. Next, with Lagrangian tracking of aerosols as passive tracers around such sedimenting hydrometeors we quantify the residence time and supersaturation experienced by individual aerosols as a function of the governing parameters.
Finally, we discuss how these results can help to quantify the likelihood and significance of heterogeneous wake-induced nucleation in atmospheric clouds.

\section{Model and Methods}

We numerically simulate the flow around a solid spherical hydrometeor with diameter $d_p$ and temperature $T_p$ falling in air (ambient: temperature $T_{\infty}$, relative humidity $RH_{\infty}$, density $\rho_a$ and pressure $p_\infty$) with constant velocity  $U_p$. While in general the shape of the hydrometeor will not be spherical and solid we expect this to be a good approximation. In the simulation all parameters are assumed to be constant, as the hydrometeor's and environment's properties vary slowly compared to that of the momentary flow (for more details see supporting information S2 and S3). 

In dimensionless form the incompressible Navier-Stokes (NS) equations and the one-way coupled advection-diffusion (AD) equations for temperature and water vapor density are,
\begin{eqnarray}
\nabla\cdot\vett{u}  &=& 0 , \label{eq.cont_ND} \\
\frac{\partial\vett{u}}{\partial t}+\vett{u}\cdot\nabla\vett{u} &=& -\nabla p +\frac{1}{Re} \nabla^2 \vett{u} , \label{eq.qdm_ND}\\
\frac{\partial T}{\partial t}+\vett{u}\cdot\nabla T &=& \frac{1}{Re\, Pr}\nabla^2 T \label{eq.T_ND} ,\\
\frac{\partial \rho_v}{\partial t}+\vett{u}\cdot\nabla \rho_v &=& \frac{1}{Re\, Sc}\nabla^2\rho_v. \label{eq.rhov_ND}
\end{eqnarray}
where $p$ is the dimensionless pressure $(p-p_\infty)/\rho_a U_p^2$, $Re=U_p d_p/\nu$ is the Reynolds number (with $\nu$ being the kinematic viscosity of air), $Pr=\nu/\kappa$ is the Prandtl number (with $\kappa$ being the thermal diffusivity of air), $Sc=\nu/\kappa_v$ is Schmidt number (with $\kappa_v$ being the water vapor diffusivity). Following  \citeA{Kotouc2009} and \citeA{Chouippe2020}, as a first approximation,  we neglect evaporation feedback on the momentum and particle temperature in the simulations(see S4).

We solve the  model equations with the lattice Boltzmann method (LBM) \cite{succi2011,Qian1992,Guo2002,Silva2012,Kruger2017,Tian2018} within the open-source LBM library Palabos \cite{Latt2020}. The simulation domain, with reference frame in the center of the hydrometeor, extends $[-5,20]d_p\times[-3.5,3.5]d_p\times[-3.5,3.5] d_p$ with a uniform Cartesian mesh of grid size $d_p/32$.
The surface of the hydrometeor is no-slip at zero velocity and with a constant temperature $T_p$ and water vapor density $\rho_{v,p}=\rho_{vs}(T_p)$ which is saturated at $T_p$ according to Maxwell diffusion model.

\section{Results}

We analyze the flow at different Reynolds numbers in the steady axisymmetric wake ($0\le Re\le 220$) and in the steady oblique wake ($225\le Re\le 285$) (see \citeA{Johnson1999}, and \citeA{Tomboulides2000}), with $Pr=0.71$ and $Sc=0.61$ according to the values of the atmospheric standard conditions \cite{Montgomery1947,Michaelides2006}.
In the case of a liquid hydrometeor of density $10^3\,\rm{kg\,m^{-3}}$ this  corresponds to a particle with a diameter between \SI{3e-4}{\metre} and \SI{1.03e-3}{\metre} falling with terminal velocities between \SI{1.21}{\metre\per\second} and \SI{4.03}{\metre\per\second}.
The ambient relative humidity $RH_{\infty}$ is varied from nearly saturated ($RH_{\infty}\sim100\%$) within the cloud \cite{Siebert_2017} to a highly subsaturated in the open atmosphere (see also \citeA{Prabhakaran2020}).
The supersaturation $S=RH-1=\rho_v/\rho_{vs}(T)-1$ is computed with respect to the water phase when $T>\SI{0}{\celsius}$ and with respect to the ice phase when $T\le\SI{0}{\celsius}$ by using the following empirical Equations \eqref{eq.Huang2018_water} and \eqref{eq.Huang2018_ice} for liquid and frozen hydrometeors respectively \cite{Huang2018}.
\begin{eqnarray}
\rho_{vs}(T)_{(T>\SI{0}{\celsius})}   &=&  \frac{1}{R_v (T+273.15)}\, \frac{\exp(34.494-{4924.99}/{(T+237.1)})}{(T+105)^{1.57}},
\label{eq.Huang2018_water}\\  
\rho_{vs}(T)_{(T\le\SI{0}{\celsius})} &=&  \frac{1}{R_v (T+273.15)}\, \frac{\exp(43.494-{6545.8}/{(T+278)})}{(T+868)^2}.
\label{eq.Huang2018_ice}
\end{eqnarray}

We identify each simulations with a defined nomenclature, like  for example, `LC 0 15 90'. Here the first letter indicates the hydrometeor phase (L liquid or I ice), the second letter indicates the sign of the temperature difference between the hydrometeor and the ambient (W warmer hydrometeor or  C colder hydrometeor), and the three following  numbers give the hydrometeor temperature $T_p$ (in degrees Celsius), the modulus of $T_p-T_\infty$ referred as $\Delta T$ (in degrees Celsius) and the ambient relative humidity $RH_\infty$ (in \%). Thus, `LC 0 15 90' is  a liquid hydrometeor colder than the ambient, with a surface temperature of \SI{0}{\celsius} in an ambient air with a temperature of \SI{15}{\celsius} and a relative humidity equal to 90\%.

\subsection{Supersaturation in the wake}

\begin{figure}
\includegraphics[width=\textwidth]{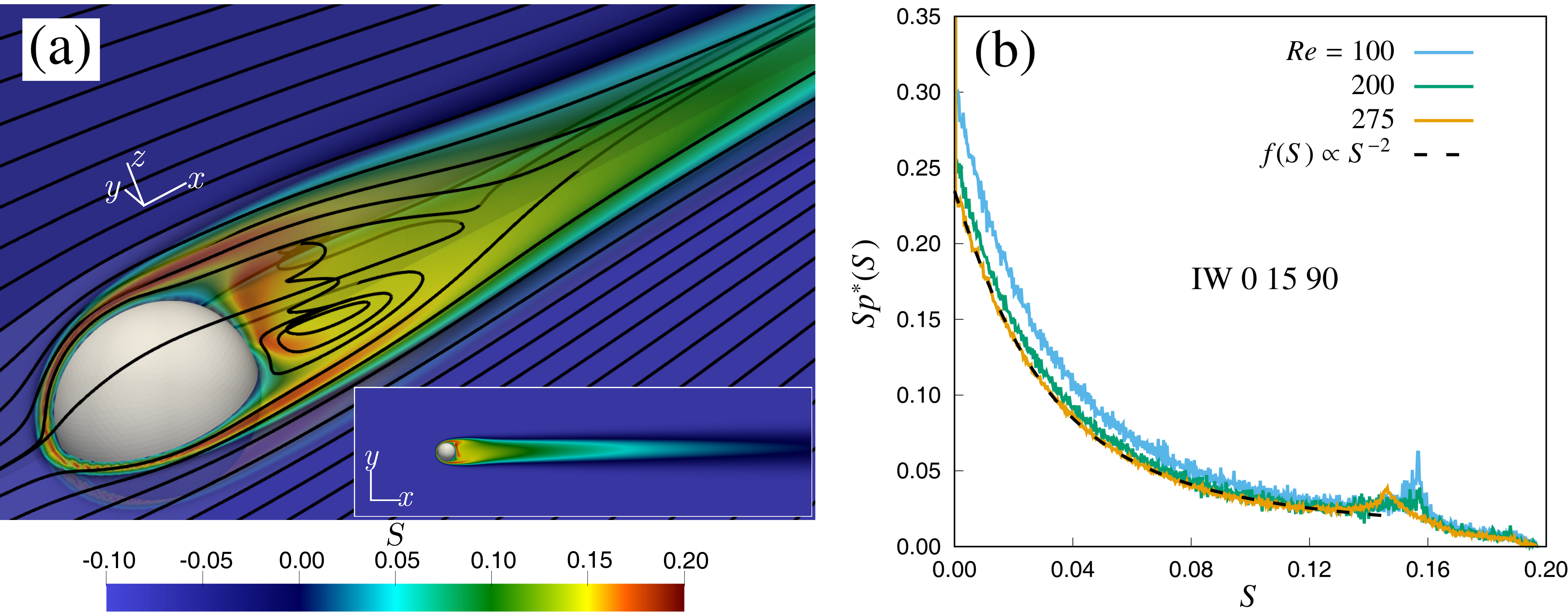}
\caption{Spatial distribution of $S$. (a) Contours of $S$ in two orthogonal central planes and complex streamlines for $Re=275$. Only the region near the hydrometeor is plotted, while the entire two dimensional domain along the orthogonal ($x,y$) plane is in the inset. (b) Normalized sample population spectrum $Sp^*(S)=Sp(S)/(\pi d_p^3/6)$ for various magnitudes of supersaturation ($S>0$) over the entire three dimensional domain for the case  `IW 0 15 90' for the $Re=100,200$ and $275$. `IW 0 15 90' represents an (I) frozen but (W) warmer hydrometeor with (0) $T_p=\SI{0}{\celsius}$ in an ambient at $T_{\infty}=\SI{-15}{\celsius}$, which gives (15) $\Delta T=\SI{15}{\celsius}$ and (90) $RH_{\infty}=90\%$ with respect to the (I) ice phase.
}
\label{fig1}
\end{figure}

Figure \ref{fig1}(a) shows an example of a visualization of the supersaturation field at $Re=275$ in an ambient relative humidity of $90\%$ with respect to ice phase for a warm hydrometeor (IW 0 15 90). High supersaturation is clearly visible in the boundary layer of the droplet and in the near wake, as well as, in the large region downstream of the hydrometeor.
In this oblique regime, some streamlines pass through the wake's vortices, a feature consistent with the results of \citeA{Johnson1999} for the oblique wake vortex structures.
The overall distribution of supersaturation in the entire three dimensional domain above a supersaturation threshold of $S_0>\SI{1e-4}{}$ is shown in Figure \ref{fig1}(b). There the supersaturation spectrum $Sp(S)$ is normalized by the hydrometeor volume and the supersaturated volume $V_{S}=\int_{S_0}^{S_{max}} Sp(S){\rm d}S$ is an integral of $Sp(S)$.
To avoid numerical round-off errors around the surface of the hydrometeor, where $S=0$, the supersaturation threshold is defined as \SI{1e-4}{}, with $S_{max}$ being the maximum supersaturation obtained in a simulation.
The statistics of the bright colored supersaturated region in Figure \ref{fig1}(a) shows the evolution $\propto S^{-2}$ in Figure \ref{fig1}(b).  The trend of $S^{-2}$ ceases around $S\ge0.13$, which is the highest magnitude of $S$ reached within the boundary layer and in the recirculating zone behind the hydrometeor in Figure \ref{fig1}(a).
$Sp(S)$ decreases slightly with increasing Reynolds number, which implies a reduction in the volume of the supersaturated region with respect to the hydrometeor volume, due to gradual thinning of the boundary layer and a correlated shrinking of the lateral extent of the wake.
Although a volumetric change in $V_{S}$ is observed with different $Re$, the magnitudes of $S_{max}$ remain almost constant for a specific thermodynamic state, independent of  $Re$.

\begin{figure}
\includegraphics[width=\textwidth]{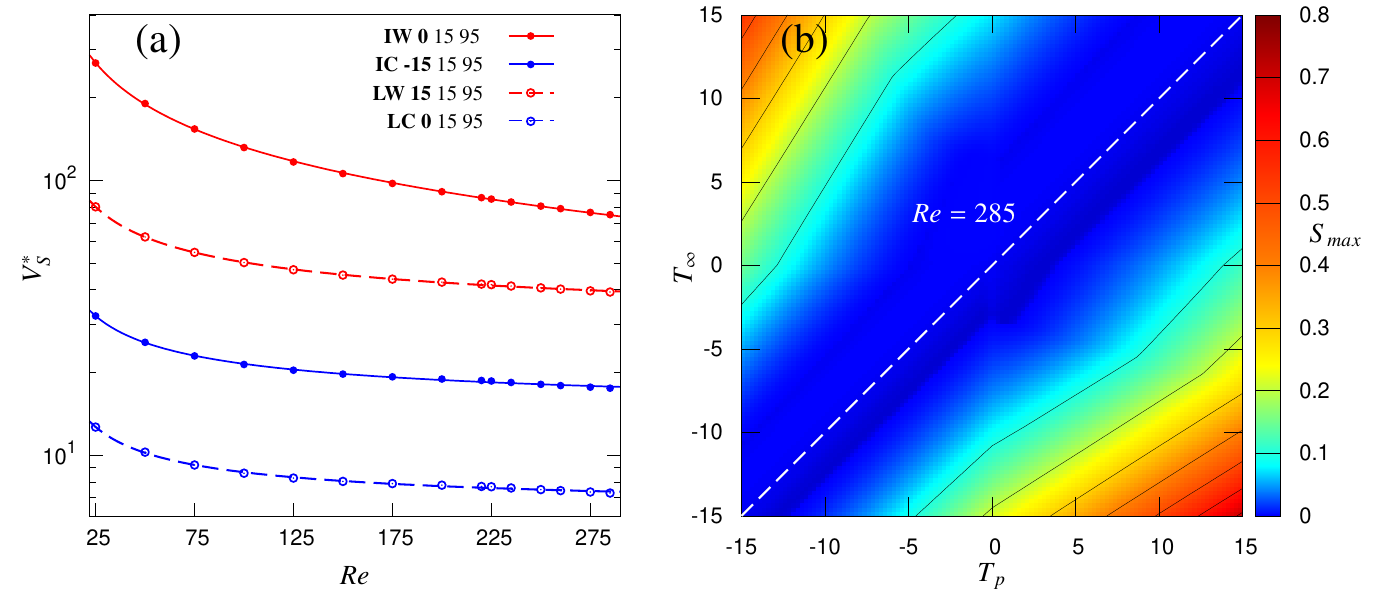}
\caption{The evolution of supersaturation in the wake. (a) The supersaturated volume $V_S^*=V_{S}/(\pi d_p^3/6)$ is plotted against different $Re$, considering both (I) frozen and (L) liquid hydrometeors with both (W) warmer or (C) colder temperature than the ambient, while maintaining $\Delta T=\SI{15}{\celsius}$ and $RH_{\infty}=95\%$.
The dots are simulation results, while the lines correspond to the $C_0(1+C_1 Re^{\alpha})$ fitting model.
Solid dots/lines represent frozen hydrometeors and the empty dots/dashed lines represent liquid hydrometeors, with red and blue color for warmer and colder hydrometeors respectively.
(b) Supersaturation maximum $S_{max}$ for different values of $T_p$ and $T_{\infty}$ varying from $-15$ to $\SI{15}{\celsius}$ for $Re=285$ keeping $RH_{\infty}=95\%$.
Black contour lines are drawn for each $0.1$ increase in $S_{max}$.
}
\label{fig2}
\end{figure}

The evolution of $V_{S}$ as a function of $Re$ and other thermodynamic parameters is shown in Figure \ref{fig2}(a) for exemplary cases presenting a temperature difference $\Delta T$ of \SI{15}{\celsius} and $RH_\infty=95\%$.
For full details on the evolution of $V_{S}$ in a whole range of $Re$, $\Delta T$, $RH_\infty$, hydrometeor phase (I or L), and warmer (W) or cooler (C) setups, see S5.
In general, a frozen hydrometeor (solid lines) 
produces a significantly larger supersaturated region than a liquid hydrometeor (dashed lines). 
This is partly due to the lower magnitude of the saturation vapor pressure in the ice phase compared to its magnitude in the liquid water phase at temperatures of $<\SI{0}{\celsius}$ (e.g., $13.7\%$ lower at $\SI{-15}{\celsius}$).
The evolution of $V_{S}$, as shown in Figure \ref{fig2}(a), with respect to the hydrometeor phase and its warmer or colder state also applies to all other $\Delta T$ and $RH_\infty$, as detailed in the supporting information. Figure \ref{fig2}(a) also shows that warmer liquid droplets, as for example, `LW 15 15 95' in $T_{\infty}=\SI{0}{\celsius}$ produce almost $2.3-2.5$ times larger $V_{S}$ than ice hydrometeors like 'IC -15 15 95'. This is generally true also for other $\Delta T$ and $RH_\infty$.
This signifies that the warmer hydrometeors produce larger $V_{S}$ than the colder ones for similar $T_{\infty}$, $\Delta T$ and $RH_{\infty}$.
This phenomenon can be further explained by analytically solving the normalized $T$ and $\rho_v$ equations (see S6) for $Re\sim0$, where warmer liquid droplets like  `LW 15 15 $RH_{\infty}$' also produce larger $V_{S}$ than the colder frozen hydrometeors as `IC -15 15 $RH_{\infty}$' for various $RH_{\infty}$ conditions.
It is further observed that a minimum of $\Delta T=4-\SI{10}{\celsius}$ is necessary to produce $V_{S}\sim \mathcal{O}(1) \times \pi d_p^3/6$, which are merely thin supersaturated boundary layers around the hydrometeor.
For hydrometeors that are colder than the ambient, $\Delta T$ needs to be at least $6-\SI{12}{\celsius}$ to produce a similar volume of $V_{S}$.

In all cases, the supersatured volume can be fitted by the following scaling function 
(goodness of fit $99.8\%$) for the whole range of the Reynolds number, despite the change in the wake structure  around $Re=220$.
\begin{eqnarray}
V_{S}=C_0(1+C_1 Re^{\alpha}) \label{eq.VS}
\end{eqnarray}
The fitting coefficient $C_0$ represents an asymptotic value, which depends on the thermodynamic parameters of the ambient and the hydrometeor , i.e., $\Delta T$, $RH_{\infty}$, (I) ice or (L) liquid, (W) warm or (C) colder temperature than the ambient.
The coefficient $C_1$ and the exponent $\alpha$ show a minor sensitivity to the thermodynamic parameters, as, $C_1$ is between $10-13$ and $\alpha$ is $-0.63\pm0.02$ for our simulations.
The data only deviates significantly when the supersaturated region is not completely within the computational domain (e.g., the case of warmer ice hydrometeors at higher Reynolds number and in almost saturated ambient) and we thus consider this a numerical artifact. 
We observed that the  $Re^{-0.63}$ scaling of the supersaturated volume closely follows the scaling of the drag coefficient with the Reynolds number in the investigated range of $Re$ \cite{Clift1978}. Thus the decrease in $V_{S}$ follows the dynamics of the wake, as also Figure \ref{fig1}(a) suggests. This aspect requires, however,  further quantitative investigation.

Figure \ref{fig2}(b) shows the development of the maximum supersaturation $S_{max}$ over a wide range of hydrometeor temperature $T_p$ and ambient temperature $T_{\infty}$ at a fixed Reynolds number $Re=285$ and an ambient relative humidity $RH_{\infty}=95\%$ for both (I) frozen and (L) liquid hydrometeors with both (W) warmer or (C) colder temperature than the ambient.
The diagonal in white dashed line corresponds to $T_p=T_\infty$ and divides the plane into the colder hydrometeor case (top left) and the warmer hydrometeor case (bottom right). The temperature difference $\Delta T$ plays a crucial role, since $S_{max}$ increases almost exponentially with it at constant $RH_{\infty}$.
Similar to $V_{S}$, warmer hydrometeors generally produce a higher supersaturation maximum than colder hydrometeors at the same $\Delta T$, regardless of their frozen or liquid state.
The only exception happens in a nearly saturated ambient at $T_{\infty}=\SI{0}{\celsius}$, because the warmer hydrometeor is a liquid one while the colder one is frozen.
In addition, $S_{max}$ evolves almost independently of $Re$ for various thermodynamic conditions. For details see S7.

\subsection{Residence time of aerosols in the wake}

Atmospheric aerosols, which can be activated as CCN or INP, behave as passive tracers due to their negligible Stokes number. To understand the possible role of the supersaturated hydrometeor wake on the aerosol activation, we have analyzed the trajectories of passive tracers injected upstream of the hydrometeor.
Since only tracers starting their motion near the center line $y=z=0$ can enter the supersaturated regions, two injection patterns are used: a coarse pattern where $2601$ tracers are injected uniformly over an area of [$1.5d_p\times1.5d_p$] and a fine pattern where $1681$ tracers are injected uniformly over an area of [$0.2d_p\times0.2d_p$] in the inlet around the hydrometeor center line.
An adaptive Runge-Kutta 4-5 method is used for time integration of the trajectories.
Velocity, temperature and vapor density at the tracer position are obtained by tri-linear interpolation.

\begin{figure}
\includegraphics[width=\textwidth]{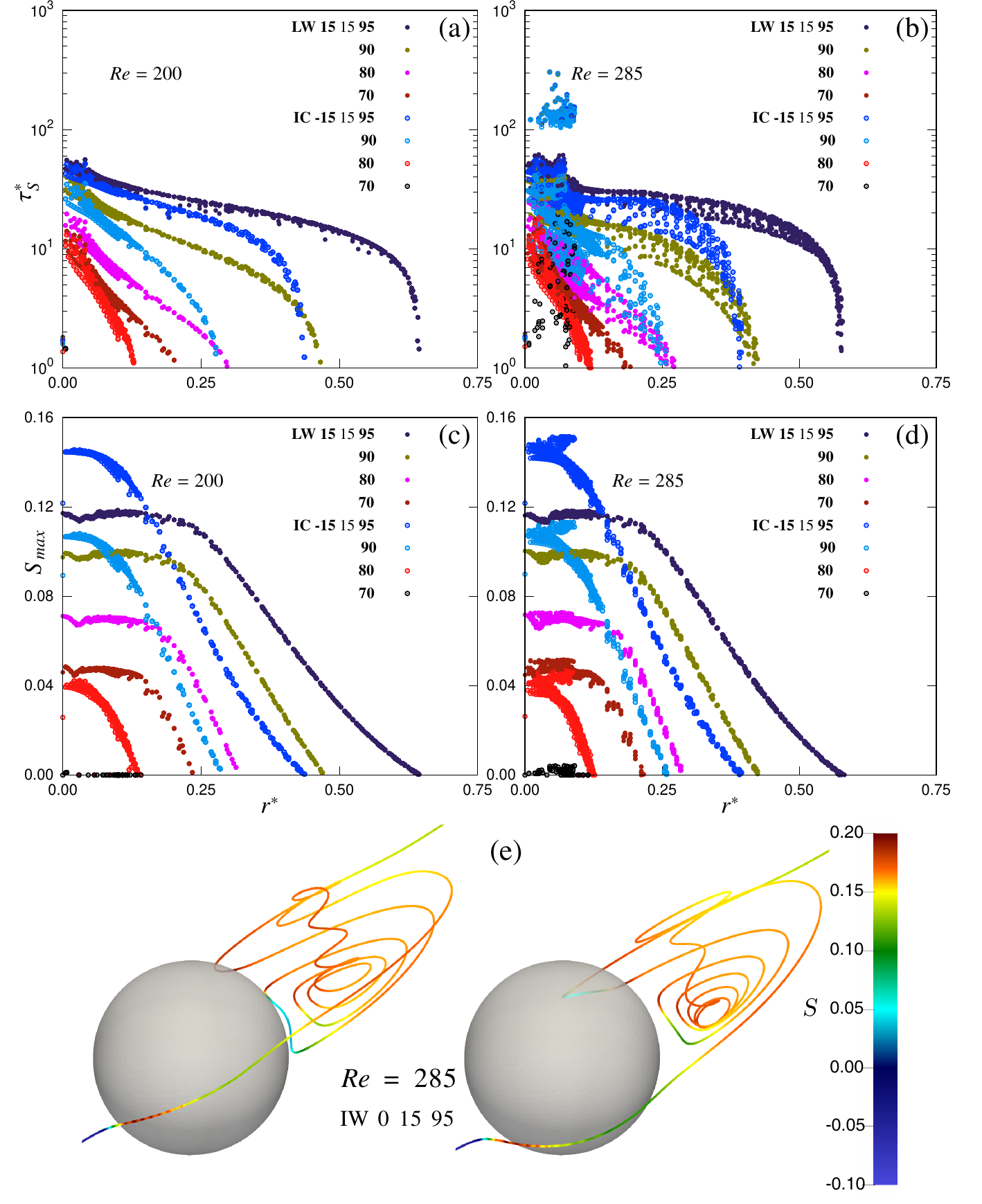}
\caption{The residence time $\tau_S^*=\tau_{S}/(d_p/U_p)$ of a tracer within the supersaturated zone and the maximum supersaturation $S_{max}$ experienced by a tracer are plotted as a function of the initial radial distance $r^*=r/d_p$ from the hydrometeor center line ($y=z=0$).
The evolution of $\tau_S^*$ within the $S>\SI{1e-4}{}$ zone is plotted for $Re=200$ in (a) and for $Re=285$ in (b), while $S_{max}$ is plotted for $Re=200$ in (c) and for $Re=285$ in (d).
Various hydrometeor phases and $RH_{\infty}$ conditions are considered keeping $\Delta T=\SI{15}{\celsius}$ and $T_{\infty}=\SI{0}{\celsius}$.
Solid and empty dots represent the liquid warm and frozen cold hydrometeor conditions, respectively.
In (e), two example tracer trajectories for $Re=285$ are shown, colored according to the instantaneous $S$ it experiences in an `IW 0 15 95' setup, resulting in $\tau_S^*=\SI{151}{\second}$ (left) and $\tau_S^*=\SI{145.4}{\second}$ (right).
}
\label{fig3}
\end{figure}

The possibility of an aerosol being activated as a CCN or an INP depends both on the instantaneous supersaturation it experiences and on the time it spends in highly supersaturated regions (residence time), so that it reaches a critical size that prevents its complete evaporation/sublimation according to the K\"ohler curve \cite{Seinfeld2006}.
In Figure \ref{fig3} we therefore plot the residence time $\tau_S$ that a tracer spends within the supersaturated wake in panels (a) and (b), and $S_{max}$ that it sees in (c) and (d) as a function of the initial radial distance $r$ of the tracer from the hydrometeor center line for axisymmetric ($Re=200$) and oblique ($Re=285$) wakes, respectively.
The different structure of the wake creates  clearly visible differences in the supersaturation experienced by the tracers.
The tracers, which stay for the longer time in the supersaturated region of axisymmetric $Re=200$ wake, are introduced near the center line as shown in Figure \ref{fig3}(a), so that they move through the supersaturated boundary layer and along the border of the wake.
However, no tracers could enter the closed recirculating region, resulting $\tau_S$ at most in the order of $10^1 d_p/U_p$ for $Re=200$.

In the oblique wake regime of $Re=285$, shown exemplary in Figure \ref{fig3}(b), tracers injected far from the axis show no significant qualitative difference in $\tau_S$ and they experience lower $S_{max}$ in Figure \ref{fig3}(d) for a short time.
However,  `lucky tracers' injected near the center line can enter the near wake vortical region and therefore remain trapped in the supersaturated recirculating zone for a longer time before moving downstream.
This increases $\tau _S$ by a factor between 2.5 to 9 with respect to the bulk of the tracers injected from the same radial distance in the symmetric or oblique wake regimes.
We quantify the extent of the injection region of lucky tracers with $\tau_S\ge10^2 d_p/U_p$, which is confined to a radial distance of $r/d_p\le0.09$.
The `capture efficiency' $E$, which is defined as the ratio between the total frontal area $A_F$ of the tracers with $\tau_S\ge 10^2 d_p/U_p$ and the frontal area of the hydrometeor $\pi d_p^2/4$, is about $5\times10^{-3}$ for $Re=285$, while it is almost zero in the steady axisymmetric regime.
The scatter in Figure \ref{fig3}(b) for $Re=285$, which produces petal-like patterns at low $r/d_p$, is due to the lack of axial symmetry in the oblique wake regime.
The larger extent of the supersaturated region generated by a warmer hydrometeor (solid dots) compared to a colder hydrometeor (empty dots) for the same $\Delta T$  and $RH_{\infty}$ is also visible in Figure \ref{fig3}.
This is evident from the lower decay of $\tau_S$ and $S_{max}$ with $r/d_p$ for warmer hydrometeors.


The mechanism allowing long residence times in the case of an oblique wake can be inferred from Figure \ref{fig3}(e), which shows two sample tracer trajectories with $r/d_p=0.078$ and $0.066$, respectively, each of which enter the vortical oblique wake region at $Re=285$.
The colors of the trajectories represent the instantaneous supersaturation the tracers experience.
Such lucky tracers, introduced very near the hydrometeor center line, experience a sudden maximum of supersaturation
$S\sim20\%$, for a short time as they move through the boundary layer on the front of the sphere.
Then the supersaturation gradually decreases along the trajectory to about $10\%$.
Later, when the tracer is entrained within the recirculating oblique wake zone, it experiences higher supersaturation again, but for a longer time due to the low velocity and complex three dimensional flow structures of this region.
However, such entrainment phenomenon is only observed when the wake loses its symmetry, i.e. in the oblique wake regime from $Re=225$ in our simulations.

\section{Implications for the Nucleation in Clouds}

The extent of the supersaturated volume, the maximum supersaturation and the residence time of an aerosol in the supersaturated wake of precipitating hydrometeors provide important insights on aerosol activation in the atmosphere.
For the aerosol entrainment in the wake, the precipitating hydrometeor has to generate an oblique wake, which occurs for a precipitating spherical raindrop when the diameter is at least \SI{1}{\milli\metre}.
Since raindrops exceeding a diameter of $2$ to \SI{3}{\milli\metre} are very rare and occur mostly in thunderstorms \cite{Pruppacher2010}, and also to satisfy the need for higher temperature difference; it is evident that wake-induced supersaturation can happen mainly in deep convective clouds with  fully glaciated, mixed phased as well as various liquid phase hydrometeors due to a large temperature variation \cite{Yuan2010}.
From the results of the previous section, the entertainment rate of  `lucky aerosols', which enter per unit time into the frontal capture area $A_F$ of a hydrometeor and thus experience a long residence time inside the supersaturated wake, is estimated as
\begin{eqnarray}
N=N_a U_p A_F=N_a U_p E \pi d_p^2/4 .\nonumber
\end{eqnarray}
\indent Here $E$ is the capture efficacy, which is about $5\times10^{-3}$ for $Re=285$ and almost zero in the steady axisymmetric regime. $N_a$ is the typical aerosol concentration, which varies from $\mathcal{O}(10^8)$ to $\mathcal{O}(10^9)\SI{}{\per\cubic\metre}$ within the continental clouds, and from $\mathcal{O}(10^7)$ to $\mathcal{O}(10^8)\SI{}{\per\cubic\metre}$ within the remote marine clouds \cite{Pruppacher2010}.
Therefore, $\mathcal{O}(10^0)\SI{}{\per\second}\le N \le \mathcal{O}(10^1)\SI{}{\per\second}$ aerosols in continental clouds and $\mathcal{O}(10^{-1})\SI{}{\per\second}\le N \le \mathcal{O}(10^0)\SI{}{\per\second}$ aerosols in remote marine clouds experience a higher residence time and higher supersaturation in the wake when a raindrop of at least \SI{1}{\mm} diameter settles at its terminal velocity ($Re\simeq285$).
Measurements of the number density of raindrops above \SI{1}{\milli\meter} show a wide variability, which can be estimated to be in the range of $\mathcal{O}(10^1)-\mathcal{O}(10^2)$ drops per cubic metre \cite<e.g.,>[and others]{Waldvogel1974,Adirosi2016}.
This leads to an entrainment rate of aerosols in the drop wakes between $\mathcal{O}(10^0)$ and $\mathcal{O}(10^3)\SI{}{\per\cubic\meter\per\second}$.
Since the capture efficiency $E$ increases with the Reynolds number (we use the $E$ of $Re=285$), this could be considered a conservative estimate. 



The critical supersaturation required for activation of aerosols is achieved by solving the K\"ohler equation for its chemical compositions and size \cite <e.g.,>[and others]{Seinfeld2006,McFiggans2006,Lohmann2015}.
Since critical supersaturation needed for the heterogeneous nucleation of common atmospheric aerosols rarely exceed $1-2\%$ in a uniform environment, we may estimate the aerosol growth (see S8) during its residence time within the supersaturated wake by considering the average supersaturation, which is much higher than $2\%$ for a temperature difference of $\SI{15}{\celsius}$ between the hydrometeor and the ambient.
Such estimation shows that inside such a supersaturated wake, an aerosol can grow well above its critical radius and therefore be activated.
During a convective precipitation process of typically 20 minutes, $\mathcal{O}(10^3)-\mathcal{O}(10^6)\SI{}{\per\cubic\meter}$ new aerosols can therefore be activated in the wake of the precipitating hydrometeors, which replenish the activated particle concentration in clouds that typically vary in $\mathcal{O}(10^8)-\mathcal{O}(10^9)\SI{}{\per\cubic\meter}$ \cite{Rosenfeld2016}.
This rate of secondary activation is well comparable with the experiment of \citeA{Mossop1976} on secondary ice production during the growth of a graupel by rimming splintering and with the in cloud measurements of secondary ice particles by \citeA{Heymsfield2014}.
However, for an explicit quantification of wake-induced nucleation, a detailed micro-physical study is required taking into account the full details of the changing atmospheric conditions and the particle evolution while falling through the convective cloud. In addition, the effects of other influencing factors, such as cloud free stream turbulence \cite{Bagchi2008}, strong convective motions like central updraft or entrainment induced mixing \cite <e.g.,>[and others]{Grabowski2013,Nair2020,Bhowmick2019}, or strong downdraft during precipitation \cite{Wang2016} may further influence this nucleation and activation rate, which needs to be carefully investigated.

\section{Summary and Concluding Remarks}

In this letter a detailed analysis of the supersaturation field and aersol activation around a spherical hydrometeor, which settles at its terminal velocity, for different atmospheric conditions is presented.
The Navier-Stokes equation for the flow velocity and the one-way coupled advection-diffusion equations for temperature and density of water vapor are solved with the lattice Boltzmann method.
The supersaturated volume $V_S$ in the wake of steady axisymmetric regime  ($Re\le220$) and oblique regime ($225\le Re\le 285$) shows a $Re^{-0.63}$ decrease for the same thermodynamic conditions.
Whereas, $V_S$ is very sensitive to the temperature difference $\Delta T$ between the hydrometeor and the ambient and its relative humidity condition $RH_{\infty}$, so that $V_{S}$ at constant $\Delta T$ increases as $RH_{\infty}$ increases, which means that a small amount of vapor diffusion from a warmer hydrometeor or cooling by a colder hydrometeor can easily supersaturate an almost saturated wake volume.
However, when $RH_{\infty}$ is fixed, $\Delta T$ plays a crucial role in $V_{S}$, since without an adequate $\Delta T$ a negligible supersaturated volume is generated.
In addition, persistently warmer hydrometeors than the ambient produced larger $V_{S}$ than the colder ones.
The supersaturation maximum $S_{max}$ behaves qualitatively similar to $V_{S}$.

Lagrangian tracking of aerosols as passive tracers shows  how the complex flow pattern of the oblique wake allows some lucky aerosols to be entrained within the recirculating wake, resulting in a higher residence time within the highly supersaturated vortical zone.
Importantly, We found that such a long residence time within the highly supersaturated wake not only exposes the aerosols to a higher level of supersaturation compared to its nucleation barrier, but also provides enough time for the growth by deposition of water vapor to exceed its critical size, and therefore to be activated as a CCN or INP.
The frontal area of these lucky tracers entering the vortical but highly supersaturated oblique wake has a capture efficiency of $\sim5\times10^{-3}$ with respect to the hydrometeor frontal area at $Re=285$.
Our analysis shows that wake-induced nucleation of aerosols during a convective precipitation of $20$ minutes can generate $\mathcal{O}(10^3)-\mathcal{O}(10^6)\SI{}{\per\cubic\meter}$ new CCNs and INPs, which is in order of magnitude comparable to other secondary ice production mechanisms, and thus contribute to the life cycle of clouds.

\acknowledgments
This research was funded by the Marie - Sk\l odowska Curie Actions (MSCA) under the European Union's Horizon 2020 research and innovation programme (grant agreement no. 675675), and an extension to programme COMPLETE by Department of Applied Science and Technology, Politecnico di Torino.
Scientific activities are carried out in Max Planck Institute for Dynamics and Self-Organization (MPIDS) and computational resources from HPC@MPIDS are gratefully acknowledged.
First author wishes to acknowledge Giuliana Donini, Guido Saracco, Mario Trigiante and Paolo Fino for support.


%
%

\bibliography{GRL2020_Bhowmick}

%
%
%
%
%


\end{document}